\documentclass{optica-article}

\journal{opticajournal} % for journals or Optica Open

\articletype{Research Article}

\usepackage{lineno}
\begin{document}

\title{Preparation of multiphoton high-dimensional GHZ states}

\author{Wen-Bo Xing,\authormark{1,2} Xiao-Min Hu,\authormark{1,2,4} Yu Guo,\authormark{1,2} Bi-Heng Liu,\authormark{1,2,3,*} Chuan-Feng Li,\authormark{1,2,3,5} and Guang-Can Guo\authormark{1,2,3}}

\address{\authormark{1}CAS Key Laboratory of Quantum Information, University of Science and Technology of China, Hefei, 230026, China\\
\authormark{2}CAS Center For Excellence in Quantum Information and Quantum Physics, University of Science and Technology of China, Hefei, 230026, China\\
\authormark{3}Hefei National Laboratory, University of Science and Technology of China, Hefei, 230088, China \\
\authormark{4}huxm@ustc.edu.cn \\
\authormark{5}cfli@ustc.edu.cn \\

\authormark{*}bhliu@ustc.edu.cn\\} %% email address is required; see note below about the corresponding author designation

% use {asbstract*} to suppress the copyright line. The copyright information will be added to production

\begin{abstract*} 
The physics associated with multipartite high-dimensional entanglement is different from that of multipartite two-dimensional entanglement. Therefore, preparing multipartite high-dimensional entanglements with linear optics is challenging. This study proposes a preparation protocol of multiphoton GHZ state with arbitrary dimensions for optical systems. Auxiliary entanglements realize a high-dimensional entanglement gate to connect the high-dimensional entangled pairs to a multipartite high-dimensional GHZ state. Specifically, we use the path degrees of freedom of photons to prepare a four-partite, three-dimensional GHZ state. Our method can be extended to other degrees of freedom to generate arbitrary GHZ entanglements in any dimension.
\end{abstract*}

%%%%%%%%%%%%%%%%%%%%%%%%%%  body  %%%%%%%%%%%%%%%%%%%%%%%%%%
\section{Introduction}
Multipartite entangled states are used to test the basic problems of quantum mechanics~\cite{hiddingvarious}. In addition, they find application in quantum computing~\cite{quantumComputational,OneWay,Nezami2020} and quantum sensing~\cite{liu2021distributed}. Multipartite entanglement is the core concept in many quantum processes, such as quantum teleportation~\cite{Bennett1993,hu2020teleportation}, dense coding~\cite{Bennett1992,guo2019advances}, and entanglement-based quantum key distribution~\cite{Scarani2009}. Compared to two-partite, two-dimensional (2D) systems, multipartite high-dimensional systems offer superior performance in fundamental research on quantum information and quantum computing.

One special multipartite entangled state is the Greenberger--Horne--Zeilinger (GHZ) state~\cite{GHZ}, $|\text{GHZ}\rangle=(|000\rangle+|111\rangle)/\sqrt{2}$. The generalized GHZ state is an entangled quantum state comprising $n>2$ subsystems. If the dimension of each system is $d$, meaning that the local Hilbert space is isomorphic to $\mathbb {C}^d$, then the total Hilbert space of the $n$-partite system is $\mathcal{H}=(\mathbb {C} ^{d})^{\otimes n}$. This GHZ state is also called the $n$-partite qudit GHZ state:
\begin{equation}
|\text{GHZ}\rangle^d_n=\frac{1}{\sqrt{d}}\left(\sum^{d-1}_{i=0}{|i\rangle^{\otimes n}}\right),\ \ (d>2,n\geq2).
    \label{dnghz}
\end{equation}
One of the remarkable properties of the GHZ state is maximizing the entanglement monotones~\cite{vidal2000entanglement,chitambar2019quantum}; therefore, it is called maximally entangled in the multipartite sense.

GHZ states find a wide range of applications. The entanglement between multiple parties is an essential feature of quantum secret sharing~\cite{secretsharing} and computational protocols~\cite{byzantineagreement}. The core support in completing these quantum protocols is the preparation of GHZ states, which has been well established in many experimental platforms, such as nuclear spins of a molecule~\cite{Negrevergne2006}, trapped ions~\cite{Haffner2005,Kaufmann2017}, nitrogen-vacancy centers in diamond~\cite{Neumann2008}, superconducting circuits~\cite{Majer2007,DiCarlo2010,Neeley2010}, silicon~\cite{Takeda2021}, and photons~\cite{Six-Photon,Eight-Photon,Ten-Photon,12-Photon}. However, these studies only prepared 2D GHZ states.

Photons are the most successful physical systems for observing multipartite entanglements ~\cite{Pan2012}. Therefore, many research groups have focused on optical systems to establish multipartite high-dimensional entanglements. Preliminary attempts in Refs. ~\cite{Malik2016a,Hu2020} for high-dimensional multiphoton produced asymmetric multipartite entanglement. To prepare high-dimensional GHZ states ($d>2$), a method called "path identity" was proposed~\cite{PathIdentity,Krenn2017b}, which was extended on the Zou--Wang--Mandel experiments~\cite{Zou1991} using a computer program~\cite{Krenn2016}. Recently, this method has been successfully used to prepare GHZ states $|\text{GHZ}\rangle^3_4$~\cite{Bao2023}.

However, "path identity" is insufficient for preparing arbitrary multiphoton high-dimensional GHZ states since "path identity" can only create arbitrarily large 2D and four-photon 3D GHZ states using graph theoretical methods till now~\cite{Krenn2017b}. Stimulated by reference~\cite{hu2020teleportation}, we proposed a method for preparing multiphoton high-dimensional GHZ states on optical systems. All the unwanted multiphoton terms can be canceled by polarizing beam splitters and auxiliary entangled photons to achieve a multiphoton high-dimensional GHZ state.

\section{Preparation of multiphoton GHZ state}

In any multiphoton entanglement preparation process, the photons from different two-partite entanglement sources are made indistinguishable (e.g., polarization, spatial mode, temporal mode, and frequency) by replacing the temporal determinant of the coherent window with ultrashort-pulsed optical pumping~\cite{zukowski1995entangling}. This significantly reduces the coherence time requirements of the interfering photons and simplifies the establishment of multiphoton 2D entanglement.

 \subsection{Four-photon GHZ state}
\begin{figure}[htbp!]
\centering
    \includegraphics [width=0.5\textwidth]{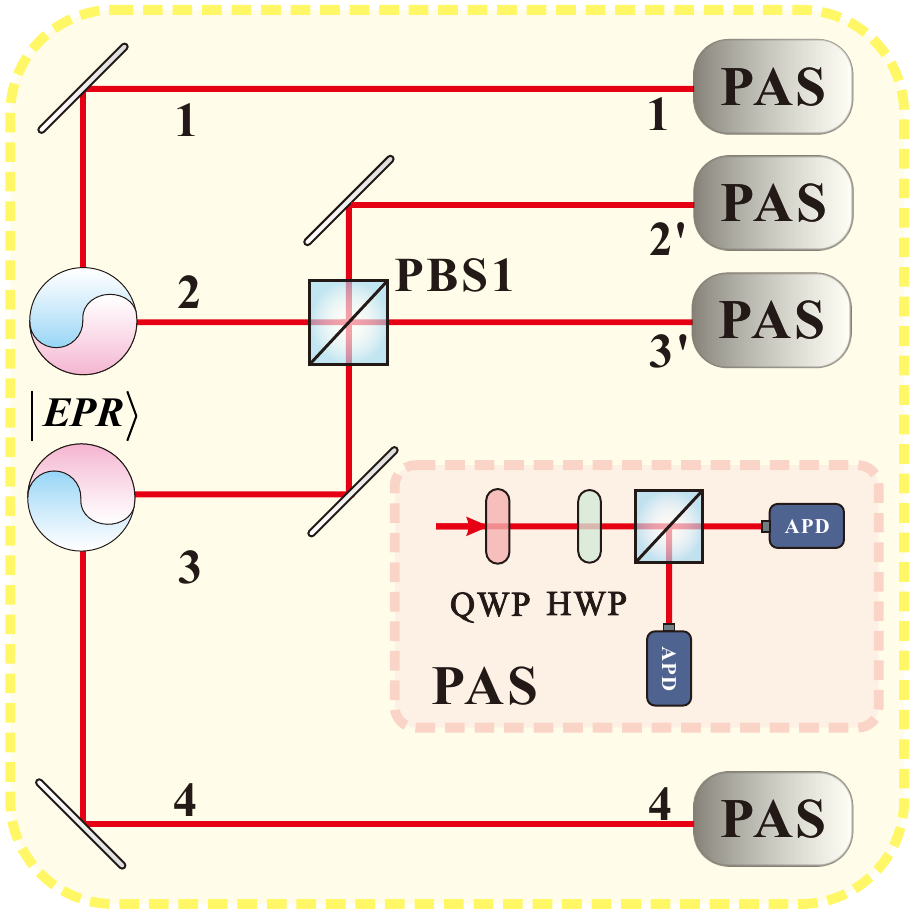}
		\caption{\textbf{experimental setup of $|\text{GHZ}\rangle^2_4$:} Two partite entanglement sources generate two EPR pairs to form the input state $|\psi^{in}\rangle=|\text{EPR}\rangle_ {12} \otimes |\text{EPR}\rangle_{34}$. PBS1 post-selects photons $3$ and $4$ with the same polarization as follows: $(O_{PBS1}=|HH\rangle\langle HH|+|VV\rangle\langle VV|)$. This operation eliminates the distinguishability of the polarization of photons $3$ and $4$. This device can be used as a unit of the multiphoton entanglement source. We can get an n-photon entanglement source if we repeat the facility of this device; PAS: polarization analysis system; QWP: quarter-wave plate; HWP: half-wave plate; PBS: polarizing beam splitter; APD: avalanche photodiode.}
  \label{4photonsetup}
\end{figure}

 We introduce the unit of this multiphoton entanglement source, which is a post-selected four-photon entanglement source. The four-photon preparation device is shown in Fig.~\ref{4photonsetup}; each Einstein-Podolsky-Rosen (EPR) source produces an entangled state.
\begin{equation}
	|\text{EPR}\rangle =\frac{1}{\sqrt{2}}\left(|HH\rangle+|VV\rangle\right),
	\label{EPR}
\end{equation}
where $|H\rangle$ and $|V\rangle$ represent the photon’s horizontal and vertical polarizations, respectively.

Therefore, our input state is
\begin{equation}
    \begin{aligned}
	|\psi^{in}\rangle^2_4 &=\frac{1}{2}\left(|HH\rangle+|VV\rangle\right)_{12}\otimes\left(|HH\rangle+|VV\rangle\right)_{34}\\
    	 &=\frac{1}{2}(|H_1H_2H_3H_4\rangle+|H_1H_2V_3V_4\rangle\\
     &\ \ \ \  +|V_1V_2H_3H_4\rangle+|V_1V_2V_3V_4\rangle).
    \end{aligned}
    \label{input1}
\end{equation}
After the photons pass through PBS1(polarizing beam splitter1 ) in  Fig.~\ref{4photonsetup}, we select the event where each exit of the PBS has one photon, indicating that the event occurs only when photons 2 and 3 with the same polarization enter PBS.
The coincidence count of the detector reveals that only the first term $|H_1H_2H_3H_4\rangle$ and fourth term $|V_1V_2V_3V_4\rangle$ remain in Eq.~\eqref{input1}.
When the Hong-Ou-Mandel (HOM) interference occurs ~\cite{HOM}, the coherent superpositioning of the two remaining terms in Eq.~\eqref{input1} yields the four-photon GHZ state
\begin{equation}
  |\text{GHZ}\rangle^2_4 = |\psi^{out}\rangle^2_4 = \frac{1}{\sqrt{2}}(|HHHH\rangle+|VVVV\rangle).
  \label{4ghz}
\end{equation}

\subsection{Multiphoton GHZ state}
The device in Fig.~\ref{4photonsetup} is scaled to construct additional photonic GHZ states.
For the GHZ state of $n$-photon, $\lceil n/2\rceil$ EPR entangled pairs are required to constitute the input state $|\psi^{in}\rangle^2_n = |\text{EPR}\rangle^{\otimes \lceil \frac{n}{2}\rceil}$. $\lceil n/2\rceil-1$ PBSs for post-selection: The processing steps in Eqs.~\eqref{input1}–\eqref{4ghz} project the input state $|\psi^{in}\rangle^2_n$ onto the target state---the $n$-photon GHZ state:
    \begin{equation}
    |\text{GHZ}\rangle^2_n = |\psi^{out}\rangle^2_n = \frac{1}{\sqrt{2}}(|H\rangle^{\otimes \lceil n/2\rceil}+|V\rangle^{\otimes \lceil n/2\rceil}),
    \label{nghz}
    \end{equation}
where $\lceil x \rceil$ ($\lfloor x \rfloor$) indicates the smallest (biggest) integer that is greater (less) than or equal to $x$. The multiphoton GHZ preparation process is an extension of the preparation process of the four-photon process. In the multiphoton preparation ($n>4$) process, we have to make adjacent EPRs interfere with each other by PBSs and post-select the multiphoton terms we wanted. Checking the HOM interference at each PBS ensures that all EPR sources are in coherent superposition.

This multiphoton extension scheme was validated through several experiments~\cite{Six-Photon,Eight-Photon,Ten-Photon,12-Photon}. The spontaneous parametric down-conversion (SPDC) method is used for preparing multiphoton entanglement sources~\cite{Zhang2021}. The real difficulty in preparing multiphoton high-dimensional GHZ states lies in the high-dimensional part generating additional multiphoton cross terms. However, these terms cannot be completely removed by PBS, thus auxiliary entanglement is required.

\section{Preparation of multiphoton high-dimensional GHZ state}
A challenging problem in constructing the high-dimensional GHZ state $|\text{GHZ}\rangle^d_n$ is how to cancel the multiphoton cross terms (MCTs) in the input state:
\begin{equation}
|\psi^{in}\rangle^d_n = \left(\sum^{d-1}_{i=0}{c_i|ii\rangle} \right)^{\otimes \lceil\frac{n}{2}\rceil}, \quad \sum{c^2_i}=1,c_i\in \mathbb{R},
\label{dninput}
\end{equation}
where $d$ and $n$ are the dimension and number of photons in the multiphoton state, respectively. The multiphoton terms that can be expressed as $|iijj\rangle$ are referred to as MCTs, where $i,j=0,1,\dots,d-1$ and $i\neq j$. In higher dimensions, to solve this problem, we introduce auxiliary entangled sources~\cite{hu2020teleportation}. We use $|\psi^{in}\rangle^3_4=[(|00\rangle+|11\rangle+|22\rangle)/\sqrt{3}]^{\otimes 2}$ to generate $|\text{GHZ}\rangle^3_4=(|0000\rangle+|1111\rangle+|2222\rangle)/\sqrt{3}$ as an example to illustrate our protocol.

\subsection{Four-Photon 3D GHZ state} 
\begin{figure}[htbp!]
    \includegraphics [width= 1\textwidth]{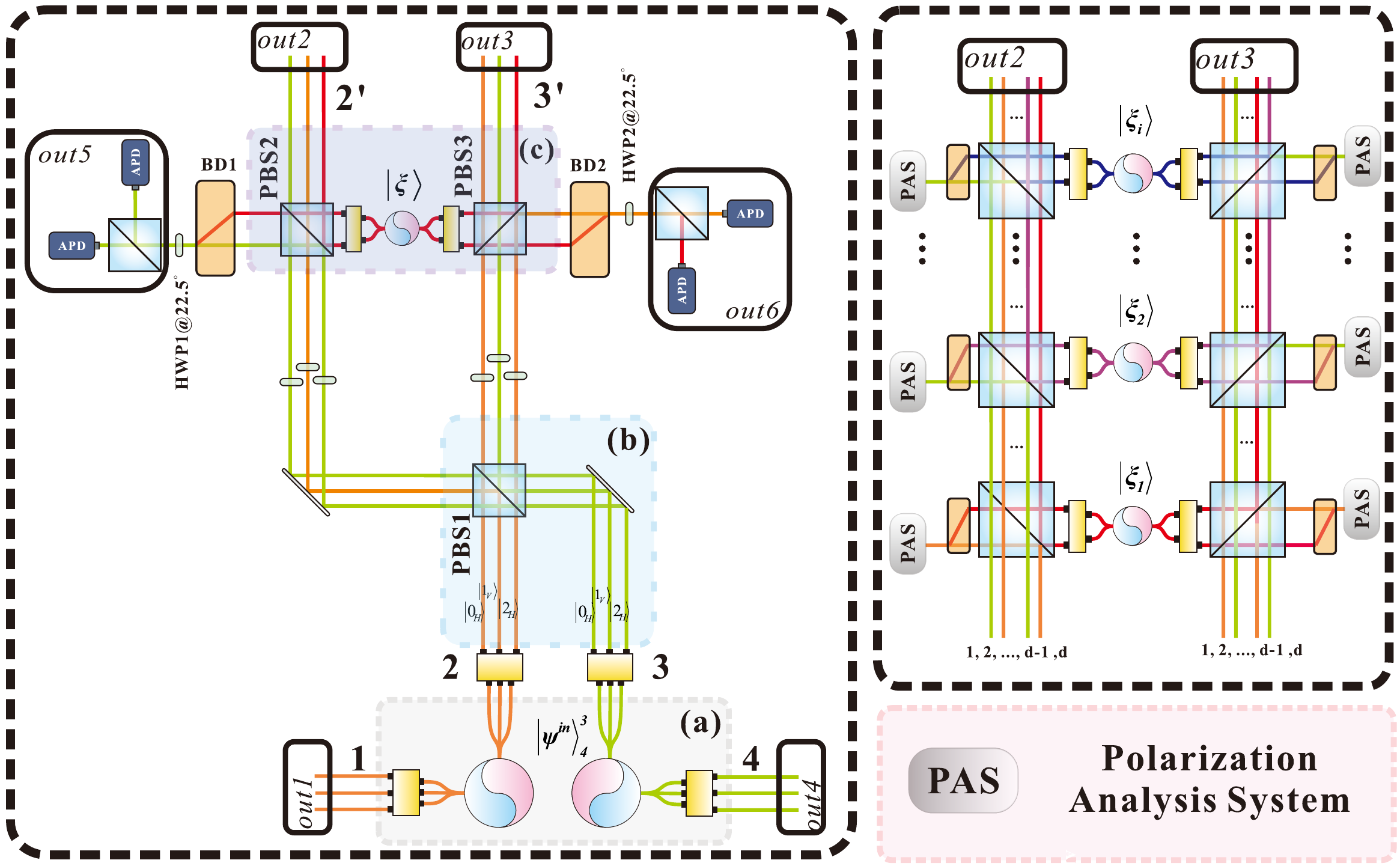}
\caption{\textbf{Diagram of four-photon high-dimensional GHZ state preparation.} The left side shows the schematic for preparing the four-photon 3D GHZ state. (a) is the input state $|\psi^{in}\rangle^3_4$. (b) Step.~i: Control the polarization of each path as shown in the diagram and post-select the state $|\psi^{in\ '}\rangle^3_4$ by PBS1. (c) Step ii: Introduce auxiliary entanglement $|\xi\rangle$ and cancel all the other MCTs after Step i. After PBS1, we only retain terms $|0 0 0 0\rangle$, $|0 0 2 2\rangle$, $|1 1 1 1\rangle$, $|2 2 0 0\rangle$, and $|2 2 2 2\rangle$. MCTs $|2 2 0 0\rangle$ and $|0 0 2 2\rangle$ are filtered out using auxiliary entanglement $|\xi\rangle$. We choose the appropriate six-fold coincidence to obtain the output state $|\text{GHZ}\rangle^3_4$ with probability $1/12$ at out1--4. The right side represents the post-selection process for any dimension. After the post-selection of PBS1, part (b) on the left side of the figure is repeated until all MCTs are removed. Different color beams indicate different entangled photons. BD: Beam displacer. PAS: Polarization
Analysis System, the structure is shown in Fig.~\ref{4photonsetup}}.
  \label{hdim}
\end{figure}

When $d = 3$, the input state can be expanded as
\begin{equation}
\begin{aligned}
    |\psi^{in}\rangle^3_4&=\left[(|00\rangle+|11\rangle+|22\rangle)/\sqrt{3}\right]^{\otimes 2}\\
    &=\frac{1}{3}(|0 0 0 0\rangle+|0 0 1 1\rangle+|0 0 2 2\rangle\\
    &\quad+|1 1 0 0\rangle+|1 1 1 1\rangle+|1 1 2 2\rangle\\
    &\quad+|2 2 0 0\rangle+|2 2 1 1\rangle+|2 2 2 2\rangle),
\end{aligned}
\label{in34}
\end{equation}
where $|i i j j \rangle$ $(i,\ j=0,1,2)$ from left to right, represents photons 1--4 in Fig.~\ref{hdim}, respectively, here we encode the systems by the path degree of freedom of photons.  

Fig.~\ref{hdim} presents the preparation idea for $|\text{GHZ}\rangle^3_4$, which is similar to $|\text{GHZ}\rangle^2_4$ described previously. We cancel all terms in the input state except $|iijj\rangle$ $(i=0\cdots2)$, which we call MCTs.

The preparation process is divided into two steps (Fig.~\ref{hdim} left).
\begin{itemize}
    \item [i)]
    Use PBS1 to remove part of the MCTs.
    \item [ii)]
    Add auxiliary entangled photons to remove remaining MCTs.
\end{itemize}

\textbf{First step}---Encode the entanglement source in the path degree of freedom. Before passing through PBS1 in Fig.~\ref{hdim}, for photons 2 and 3, if the photon is in state $|0\rangle$ or $|2\rangle$, we set the polarization to horizontal, which is noted as {$|0_H\rangle$, $|2_H\rangle$}. Similarly, when the photon is in state $|1\rangle$, we note it as $|1_V\rangle$. After the photons pass through PBS1, we choose the event that each exit of PBS1 has only one photon. Therefore, $|00_H1_V1\rangle$, $|11_V0_H0\rangle$, $|11_V2_H2\rangle$, and $|22_H1_V1\rangle$ in Eq.~\eqref{in34} are dropped. Thus, we obtain the state described as
\begin{equation}
\begin{aligned}
    |\psi^{in}\ '\rangle^3_4=\frac{1}{3}&(|0 0_H 0_H 0\rangle+|0 0_H 2_H 2\rangle+|1 1_V 1_V 1\rangle\\
    &+|2 2_H 0_H 0\rangle+|2 2_H 2_H 2\rangle ).
\end{aligned}
\label{in342}
\end{equation}
Compared with Eq.~\eqref{in34}, we obtain the state in Eqs.~\eqref{in342} with a success probability of $p_1=5/9$.

\textbf{Second step}---MCTs have two other unwanted terms, $|0 0 2 2 \rangle$ and $|2 2 0 0 \rangle$. After PBS1, we set the polarization of path $\{|0\rangle $, $|1\rangle$, $|2\rangle\}$ on $\{|0_H\rangle $, $|1_H\rangle$, $|2_V\rangle\}$ respectively, and introduce the auxiliary entanglement $|\xi\rangle=(|0_H0_H\rangle+|2_V2_V\rangle)/\sqrt{2}$ as: 
\begin{equation}\label{fz}
\begin{aligned}
    |\psi^{in}\ '\rangle^3_4\otimes |\xi\rangle&=\frac{1}{3}(|0 0_H 0_H 0\rangle+|0 0_H 2_V 2\rangle+|1 1_H 1_H 1\rangle+|2 2_V 0_H 0\rangle+|2 2_V 2_V 2\rangle)\otimes |\xi\rangle \\
    &=\frac{\sqrt{2}}{6}[|0 0_H 0_H 0 0_H 0_H\rangle+|0 0_H 0_H 0 2_V2_V\rangle+|1 1_H 1_H 1 0_H 0_H\rangle\\
    &\quad\quad+|1 1_H 1_H 1 2_V 2_V\rangle+|0 0_H 2_V 2 0_H 0_H\rangle+|0 0_H 2_V 2 2_V 2_V\rangle\\
    &\quad\quad+|2 2_V 0_H 0 0_H0_H\rangle+|2 2_V 0_H 0 2_V2_V\rangle+|2 2_V 2_V 2 0_H0_H\rangle\\
    &\quad\quad+|2 2_V 2_V 2 2_V2_V\rangle].
\end{aligned}
\end{equation}
After the photons pass through PBS2--3, we focus on path $|0\rangle, |2\rangle$, and ignore the case where two photons go to the same side. Thus we can obtain the following state with probability $p_2=3/10$:
\begin{equation}\label{p2}
\frac{\sqrt{2}}{6}(|0 0_H 0_H 0 0_H 0_H\rangle+|1 1_H 1_H 1 0_H 0_H\rangle+|2 2_V 2_V 2 2_V2_V\rangle).
\end{equation}
Finally, the photons pass through HWP1--2 and BD1--2 to obtain the following state:
\begin{equation}\label{eq:12}
\begin{aligned}
      \frac{\sqrt{2}}{6}[& \frac{1}{2}|0 0 0 0\rangle \otimes(|HH\rangle+|VV\rangle+|HV\rangle+|VH\rangle)\\
      &+\frac{1}{2}|1 1 1 1\rangle\otimes(|HH\rangle+|VV\rangle+|HV\rangle+|VH\rangle)\\
      &+\frac{1}{2}|2 2 2 2\rangle \otimes(|HH\rangle+|VV\rangle-|HV\rangle-|VH\rangle)].
\end{aligned}
\end{equation}
We rewrite the above state as:
\begin{equation}
    \begin{aligned}
      &\frac{1}{2\sqrt{3}}[\frac{1}{\sqrt{3}}(|0 0 0 0\rangle+|1 1 1 1\rangle+|2 2 2 2\rangle)]\otimes \frac{1}{\sqrt{2}}(|HH\rangle+|VV\rangle)\\
        &+\frac{1}{2\sqrt{3}}[\frac{1}{\sqrt{3}}(|0 0 0 0\rangle+|1 1 1 1\rangle-|2 2 2 2\rangle)]\otimes \frac{1}{\sqrt{2}}(|HV\rangle+|VH\rangle).
\end{aligned}
\label{eq:mutifold}
\end{equation}
We choose the six-fold coincidence count with $|HH\rangle$ or $|VV\rangle$ at out5--6 with probability $p_3=1/2$ and obtain the following state:
\begin{equation}
|\psi^{out}\rangle ^3_4=\frac{1}{2\sqrt{3}}[\frac{1}{\sqrt{3}}(|0 0 0 0\rangle+|1 1 1 1\rangle+|2 2 2 2\rangle)]=\sqrt{p_1 p_2 p_3}\ |\text{GHZ}\rangle^3_4.
\label{eq10}
\end{equation}
We can use the input state $|\psi\rangle^3_4$ from Eq.~\eqref{in34} with probability $\prod^3_{i=1}{p_i}=\frac{1}{12}$ to obtain the 4-photon 3D GHZ state $|\text{GHZ}\rangle^3_4$.

Furthermore, when we choose the six-fold coincidence count with $|HV\rangle$ or $|VH\rangle$ in out5--6, we can obtain a quantum state $1/\sqrt{3}(|0 0 0 0\rangle+|1 1 1 1\rangle-|2 2 2 2\rangle)$, which is not a GHZ state. If based on the measurement results, we apply a $\pi$ phase modulation to the photons in path 2, which can be deterministically accomplished with an electro-optic feedforward~\cite{Sciarrino2006,Prevedel2007,Ma2011,Svarc2020}, we can obtain the same GHZ state as in Eq.~\eqref{eq10}. At this point, the success probability of obtaining the 4-photon 3D GHZ state is $1/6$.

\subsection{High-dimensional GHZ state}
The protocol is scalable to arbitrary dimensions. The preparation protocol for $d\geq2$ with four-photon GHZ states $|\text{GHZ}\rangle^d_4$ in arbitrary dimensions can be generalized as follows:
\begin{enumerate}
    \item [i)]
    Prepare two 2-photon $d$-dimensional maximal entangled state: $|\phi\rangle^d=(\sum_0^{d-1}|ii\rangle)/\sqrt{d}$ which encode by the path degree of freedom, and control the polarization of each path as follows: if $i$ is odd, the polarization is set at $V$; otherwise, it is set at $H$. Then, when the photons pass through PBS1 in Fig.~\ref{hdim}, we select the events with only one photon at each exit of PBS1.\label{item1}
    \item [ii)] \label{item2}
Set the polarization of photon in the path state $|i\rangle \ (|j\rangle)$ at $|i_H\rangle \ (|j_V\rangle)$. Prepare auxiliary entangled state $|\xi\rangle=(|i_Hi_H\rangle)+|j_Vj_V\rangle)/\sqrt{2}$ $(i,j=0,\cdots d-1.\ i < j)$ to cancel MCTs  $|iijj\rangle$ and $|jjii\rangle$.
    \item[iii)]  \label{item3}
    Repeat Step.~ii until all MCTs are canceled.
\end{enumerate}
After Step.~i, the MCTs between the odd- and even-numbered paths are canceled; therefore, in Step.~ii, $i,\ j$ is either odd or even. There are $\lceil d/2 \rceil$ paths encoded as even and $\lfloor d/2 \rfloor$ encoded as odd. When we repeat Step.~ii $\operatorname{C}_{\lceil d/2\rceil}^{2}$ ( $\operatorname{C}_{\lfloor d/2\rfloor}^{2}$) times, we can cancel MCTs between odd- (even-) numbered paths. We conclude that $N^d_4=\operatorname{C}_{\lceil d/2\rceil}^{2}+\operatorname{C}_{\lfloor d/2\rfloor}^{2}=\lceil d(d-2)/4\rceil$ auxiliary entanglements are required to prepare four-photon GHZ states in $d$ dimension. It should be noted that before any post-selection by PBSs, the HOM interference on the PBS must be checked to ensure the coherent superposition of the auxiliary entangled photons and the $|\phi\rangle^d$ photons.

The post-selection in Step.~i filter out all MCTs between the odd and even paths (i.e, the MCTs which can be written as $|iijj\rangle$, $i$ for odd numbers and $j$ for even ones, or vice versa), which have $2\lceil d/2 \rceil \times\lfloor d/2 \rfloor$ terms. Thus the success probability of Step.~i in $d$ dimension is $\eta_1=1-2\lceil d/2 \rceil \times\lfloor d/2 \rfloor/d^2$. Two more MCTs will be canceled whenever Step.~ii is performed. Some multiphoton terms are retained with probability $1/2$, e.g. $|iiii\rangle$ $(i=0, 1, 2)$ in the preparation of $|\text{GHZ}\rangle^3_4$ is retained with probability $1/2$, and with the help of $\pi$ phase modulation, the probability of selecting the correct multifold coincidence is $1$. Thus, the success probability of $k$th Step.~ii is:
\begin{equation}
    \eta_2(k)=\frac{d^2 \eta_1 -2k}{2 [d^2 \eta_1-2(k-1)]}.
\end{equation}
The success probability in the preparation of $|\text{GHZ}\rangle^d_4$ is $p=\eta_1 \times \prod^{N^{d}_4}_{k=1}{\eta_2(k)}=1/2^{N^{d}_4}\times(\eta_1-2N^{d}_4/d^2)$, where $\eta_1-2N^{d}_4/d^2=1/d$, thus $p=1/2^{N^{d}_4}\times 1/d$. If $d=3$, then $p=1/6$. This is consistent with the result in Section 3.1.

\subsection{Multiphoton high-dimensional GHZ state} 
\begin{figure}[htbp!]
\centering
    \includegraphics [width=0.8\textwidth]{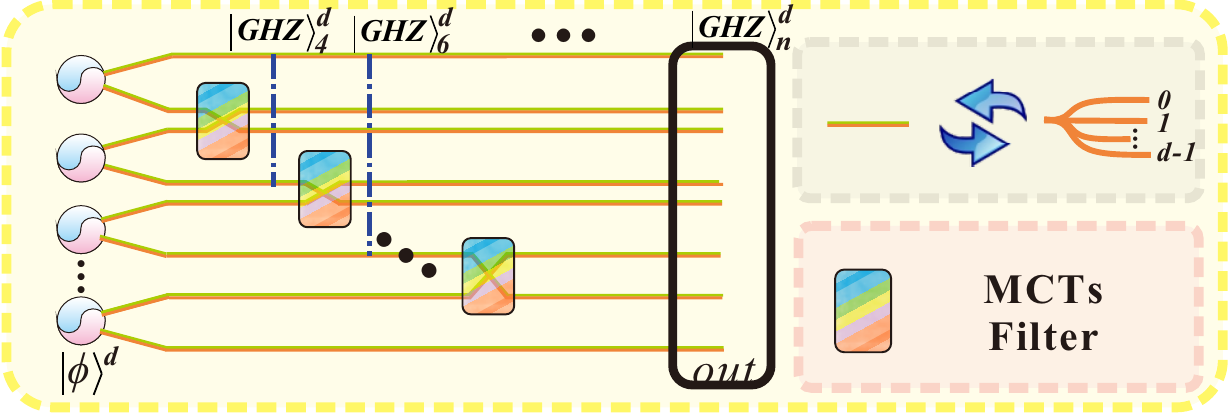}
		\caption{\textbf{ Schematic diagram of multiphoton high-dimensional GHZ state $|\text{GHZ}\rangle^d_n$ preparation :} We start from the state $|\text{GHZ}\rangle^d_4$. When two adjacent photons belonging to different $|\phi\rangle^d$ and pass through the MCTs filter, which allows filtering out the multiphoton term unrelated to the GHZ state. Then we add another $|\phi\rangle^d$ and the MCTs filter, and after a similar post-selection process, we get the state $|\text{GHZ}\rangle^d_6$. We prepare the state $|\text{GHZ}\rangle^d_n=(|0\rangle^{\otimes n}+|1\rangle^{\otimes n}+\dots +|d-1\rangle^{\otimes n})/\sqrt{d}$ by repeating the process of combining $|\phi\rangle^d$ and the MCTs filters. The MCTs filter contains two post-selection processes. The first process is post-selection using PBS (as illustrated in Fig~.\ref{hdim}(a)), while the second involves auxiliary entangled photons (as illustrated on the right side of Fig~.\ref{hdim}).}
  \label{fig:hdpsetup}
\end{figure}

 Fig.~\ref{fig:hdpsetup} shows the diagram of the preparation of $|\text{GHZ}\rangle^d_n$. 
We prepare the state $|\text{GHZ}\rangle^d_4$ by sending the adjacent photons from the two 2-photon $d$-dimensional maximal entangled state: $|\phi\rangle^d=(\sum_0^{d-1}|ii\rangle)/\sqrt{d}$ through the MCTs filter (shown as the first blue line in Fig.~\ref{fig:hdpsetup}.), which filters out multiphoton terms unrelated to the GHZ state. We proceed to prepare the entanglement source $|\phi\rangle^d$ and pass it through the MCTs filter, and obtain $|\text{GHZ}\rangle^d_n$ state. The post-selection processes by PBSs and auxiliary entanglements are included in the MCTs filter. The MCTs filter is implemented in two parts and contains all the post-selection processes. The first part includes the post-selection using PBS1, shown in Fig.~\ref{hdim}(a). The second part contains the post-selection part using several auxiliary entanglements as shown in the right of Fig.~\ref{hdim}.

The number of photons in our input state is always an even number. When we want to prepare a GHZ state with an odd number of photons, we need to perform the measurement $M^d=|m\rangle\langle m|$ for the photons of out1, where the measurement basis $|m\rangle=1/\sqrt{d}(|0\rangle+|1\rangle+\dots+|d-1\rangle)$, so that we will change with probability $1/d$ from the GHZ state of even photons to odd number. To improve the probability, we can perform a $d$-output measurement of the photon at out1. Let us take an example in three dimensions: if one of our measurements base is $M^3$, then the measurement bases of the remaining two measurement bases are: $1/\sqrt{d}(|0\rangle+e^{\frac{2\pi}{3}i}|1\rangle+e^{\frac{4\pi}{3}i}|2\rangle)$ and $1/\sqrt{d}(|0\rangle+e^{\frac{4\pi}{3}i}|1\rangle+e^{\frac{2\pi}{3}i}|2\rangle)$. We can find that this $d$-output measurement corresponds to a complete Fourier base measurement performed on the photon at out1. Each measurement output represents the photon performing a Fourier basis measurement that would make it have $e^{\frac{2\pi}{d}i}$ phases between adjacent multiphoton terms ($|iii\rangle$) of our final state. If we apply the phase modulation device from the previous section, we make it so that the photon in out1 is capable of preparing a GHZ state regardless of which measurement output it is detected by.

The multiphoton entanglement source requires $\lceil n/2\rceil$ $|\phi\rangle^d$ and $N^d_n=\lceil d(d-2)/4\rceil \times \lceil n/2-1\rceil$ auxiliary two-partite 2D entanglement sources to filter out the MCTs. When preparing a multiphoton high-dimensional GHZ state, the input state contains $d^{\lceil n/2\rceil}$ multiphoton terms, of which we preserve only $d$ of them (the probability is $1/d^{\lceil n/2\rceil-1}$). If auxiliary entanglement is introduced, the retained multiphoton terms will be kept with a probability of $1/2$ (e.g., multiphoton terms $|iiii\rangle$ will be retained by probability with $1/2$). Meanwhile, when we choose the correct coincident events, we should keep the PAS pairs measure $|HH\rangle$ or $|VV\rangle$ (if measure $|HV\rangle$ or $|VH\rangle$, we should add a $\pi$ phase modulation in path 2). So we have a probability of $1$ to obtain our target state.  As a result, the success probability of preparing the multiphoton high-dimensional GHZ state using our protocol is $1/d^{\lceil n/2\rceil-1}\times 1/2^{N^d_n}$. 
The step-by-step analysis becomes complex since the preparation procedure demands more post-selection processes. We determine the final preparation success probability by comparing the number of multiphoton terms in the output and input states, which is comparable to the preceding section's method.

\section{Conclusion}
Our protocol is feasible using current experimental techniques. The key is selecting the target state through the joint action of PBSs and auxiliary entanglements~\cite{hu2020teleportation} after multiple post-selections and HOM interferences.

The protocol is divided into two steps. The first step is the preparation of 4-photon high-dimensional states; the second step is to prepare additional $|\phi\rangle^d$ and apply MCTs filters to obtain $n$-photon ($n>4$) high-dimensional GHZ states, as shown in Fig.~\ref{fig:hdpsetup}. The process is illustrated using an example of a four-photon 3D GHZ state $|\text{GHZ}\rangle^3_4$. Finally, we derived the requirement of high-dimensional entanglement sources and auxiliary entanglement sources for preparing arbitrary multiphoton high-dimensional GHZ state $|\text{GHZ}\rangle^d_n$, and give the success probability of the process.

However, an important question remains on how to efficiently reduce auxiliary entanglement and improve the success probability. The solution will facilitate the scalability and stability of the dimensions of the experimental setup used in real experiments, and improve the feasibility of experimental protocols.

In contrast to the "path identity", our protocol successfully solves the problem of preparing arbitrary multiphoton high-dimensional GHZ states in optical systems. With the development of quantum information technology, this protocol makes many quantum protocols have better application prospects in optical systems. Although our protocol is assumed to be performed in a bulk optical system setup using path encoding, it can be applied to other degrees of freedom of photons and integrated optical system.

%%%%%%%%%%%%%%%%%%%%%%% References %%%%%%%%%%%%%%%%%%%%%%%%%
\begin{backmatter}
\bmsection{Funding}
This work was supported by the National Key Research and Development Program of China (No. 2021YFE0113100), NSFC (No. 11904357, No. 12174367, and No. 12204458), Innovation Program for Quantum Science and Technology (No. 2021ZD0301200), Fundamental Research Funds for the Central Universities, USTC Tang Scholarship, Science and Technological Fund of Anhui Province for Outstanding Youth (2008085J02), China Postdoctoral Science Foundation (2021M700138), and China Postdoctoral for Innovative Talents (BX2021289), the Shanghai Municipal Science and Technology Fundamental Project (No. 21JC1405400). This work was partially supported by the USTC Center for Micro- and Nano-scale Research and Fabrication. 

\bmsection{Disclosures} The authors declare no conflicts of interest.

\bmsection{Data availability} No data were generated or analyzed in the presented research.

\end{backmatter}

\bibliography{reflog}

\end{document}